\documentclass[twocolumn,preprintnumbers,amsmath]{revtex4}
\usepackage{graphicx}% Include figure files   twocolumn,
\usepackage{dcolumn}% Align table columns on decimal point
\usepackage{bm}% bold math
\usepackage{setspace}
\usepackage{color}
\usepackage{overpic}
\usepackage{sidecap}
\usepackage{braket}

\usepackage[english]{babel}
\draft

\begin{document}

\title{Spontaneous creation of Kibble-Zurek solitons in a Bose-Einstein condensate}

\author{Giacomo Lamporesi}
\author{Simone Donadello}
\author{Simone Serafini}
\author{Franco Dalfovo}
\author{Gabriele Ferrari}
\affiliation{INO-CNR BEC Center and Dipartimento di Fisica, Universit\`a di Trento, 38123 Povo, Italy}

\begin{abstract}
When a system crosses a second-order phase transition on a finite timescale, spontaneous symmetry breaking can cause the development of domains with independent order parameters, which then grow and approach each other creating boundary defects. This is known as Kibble-Zurek mechanism. Originally introduced in cosmology, it applies both to classical and quantum phase transitions, in a wide variety of physical systems. Here we report on the spontaneous creation of solitons in Bose-Einstein condensates via the Kibble-Zurek mechanism. We measure the power-law dependence of defects number with the quench time, and provide a check of the Kibble-Zurek scaling with the sonic horizon. These results provide a promising test bed for the determination of critical exponents in Bose-Einstein condensates.

\end{abstract}

\maketitle

\begin{figure*}[!]
\centering
\includegraphics[width=2.05\columnwidth]{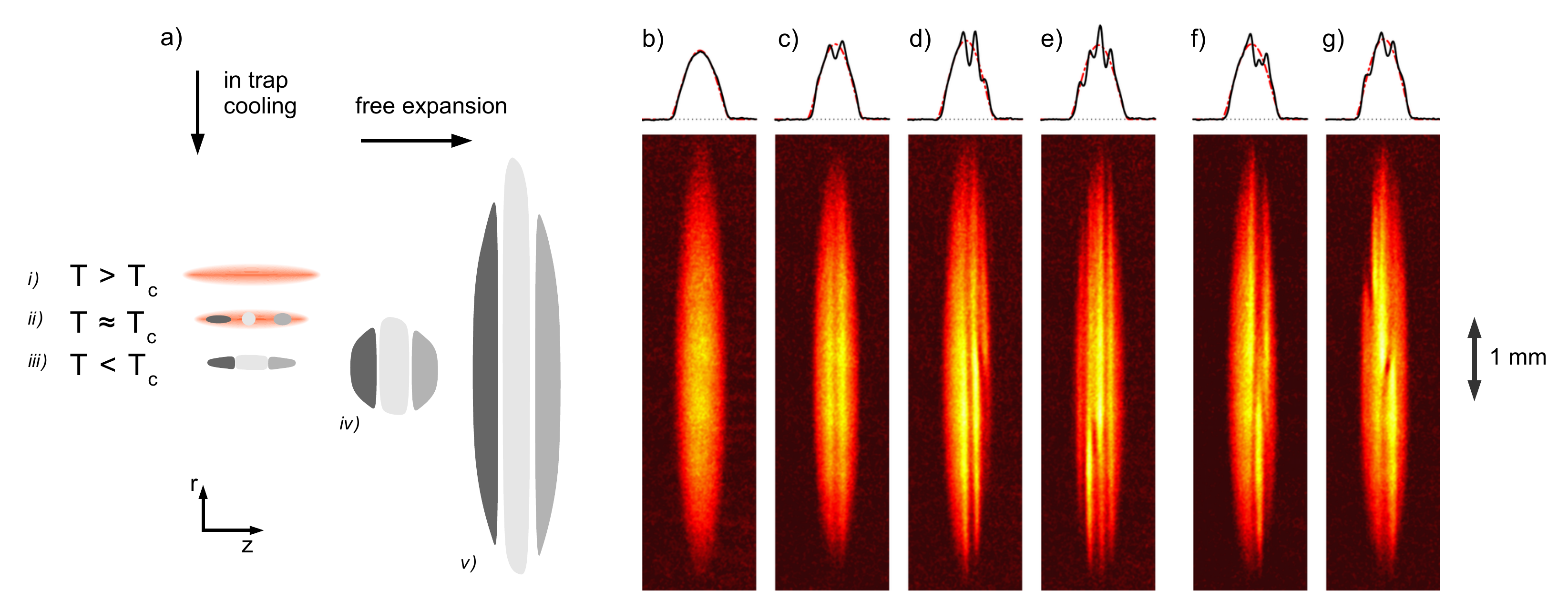}
\caption{\textbf{Solitons in an elongated BEC:} \textbf{a}, Formation after a quenched cooling on a thermal gas (\emph{i}, red) across the BEC transition, BEC is locally achieved forming several isles (\emph{ii}) each with its own phase (grey). Further cooling makes them grow and get close (\emph{iii}) forming solitons. The sample is released from the trap and let expand for 180 ms (\emph{iv}-\emph{v}) in a levitating field. \textbf{b}-\textbf{e}, Sample pictures of the BEC after expansion containing 0,1,2,3 solitons or even fancier structures with bendings and crossings (\textbf{f}-\textbf{g}). For each picture the integrated profiles of the central region (1/3 of the Thomas-Fermi diameter) are shown in black and compared to the parabolic Thomas-Fermi fit in red.}
\label{fig:solitons}
\end{figure*}

The Kibble-Zurek mechanism (KZM) describes the spontaneous formation of defects in systems that cross a second-order phase transition at finite rate \cite{Kibble76,Kibble80,Zurek85,Zurek96}. The mechanism was first proposed in the context of cosmology to explain how during the expansion of the early Universe the rapid cooling below a critical temperature induced a cosmological phase transition resulting in the formation domain structures. In fact, the KZM is ubiquitous in nature and regards both classical and quantum phase transitions \cite{Dziarmaga2002,Zurek2005}. Experimental evidences have been observed in superfluid $^4$He \cite{Hendry94,Dodd98} and $^3$He \cite{Bauerle96,Ruutu96}, in superconducting films \cite{Carmi99} and rings \cite{Carmi00,Kavoussanaki00,Monaco02,Monaco03,Monaco09} and in ion chains \cite{Ulm13,Ejtemaee2013}. Bose-Einstein condensation in trapped cold gases has been considered as an ideal platform for the KZM \cite{Zurek09,Damski10,delCampo11,Witkowska11,Sabbatini2011}; the system is extremely clean and controllable and particularly suitable for the investigation of interesting effects arising from the spatial inhomogeneities induced by the confinement. Quantized vortices produced in a pancake-shaped condensate by a fast quench across the transition temperature have been already observed \cite{Weiler08}, but their limited statistics prevented the test of the KZM scaling. The KZM has been studied across the quantum superfluid to Mott insulator transition with atomic gases trapped in optical lattices \cite{Chen11}. Here we report on the observation of solitons resulting from phase defects of the order parameter, spontaneously created in an elongated Bose-Einstein condensate (BEC) of sodium atoms. We show that the number of solitons in the final condensate grows according to a power-law as a function of the rate at which the BEC transition is crossed, consistent with the expectations of the KZM, and provide the first check of the KZM scaling with the sonic horizon. We support our observations by comparing the estimated speed of the transition front in the gas to the speed of the sonic causal horizon, showing that solitons are produced in a regime of inhomogeneous Kibble-Zurek mechanism (IKZM) \cite{delCampo11}. Our measurements can open the way to the determination of the critical exponents of the BEC transition in trapped gases, for which so far little information is available \cite{Donner07}.\\

The KZM predicts the formation of independent condensates when the system crosses the BEC transition at a sufficiently fast rate (Fig.\,\ref{fig:solitons}a \textit{i}-\textit{ii}). Further cooling and thermalization below the critical temperature causes the independent condensates to grow. In axially elongated trapping potentials neighboring condensates with different phases will approach forming solitons \cite{Zurek09} (Fig.\,\ref{fig:solitons}a \textit{iii}). We characterize this process by counting the solitons as a function of the quench time and the atom number at the transition by means of direct imaging after a ballistic expansion of the sample (Fig.\,\ref{fig:solitons}a \textit{iv}-\textit{v}). Typical density distributions after time-of-flight (TOF) are shown in Fig.\,\ref{fig:solitons}b-g. The case in panel b) corresponds to a condensate with negligible thermal component and almost in its ground state. Panel c), instead, shows a density depletion which we interpret as a soliton. More solitons are shown in the other panels, including cases where the solitonic planes are bent and/or collide as in f) and g). As opposed to artificially created solitons via phase imprinting techniques \cite{Burger98,Denschlag00,Becker08} or by exciting the superfluid with laser pulses or through collisions \cite{Chang08,Shomroni09}, our solitons spontaneously form when the BEC is created by crossing the transition temperature.

The identification of these defects as dark/grey solitons is based on several arguments: they are simultaneously observed as lines from two orthogonal directions in the radial plane, demonstrating their planar structure, mostly perpendicular to the weak axis of confinement; sometimes they exhibit a bent shape as we expect for snake oscillations \cite{Anderson2001} of soliton planes; when two of these defects overlap, they appear as solitons in a collision \cite{Carr2008}, whose individual structure is preserved except in the crossing region.  Finally their size after TOF is of the right order of magnitude.
This can be deduced by considering that the width of a soliton is of the order of the healing length $\xi = (8\pi a n)^{-1/2}$, where $a$ is the scattering length and $n$ the spatial density. One can then assume that, during the initial fast expansion of the gas in the radial direction, the healing length increases by adiabatically following the density reduction, similarly to what happens to the cores of quantized vortices in a disk-shaped condensate subject to a rapid expansion in the axial direction \cite{Dalfovo00}. As a consequence, a long expansion time allows for a better visibility and counting resolution. The expansion times we chose for imaging are indeed much longer than standard ones, thanks to an external magnetic field gradient used for levitating the gas against gravity; this is essential to reduce the optical density well below saturation and for solitons to become large enough to be clearly detected. \\

\begin{figure*}[!]
\centering
\includegraphics[width=2.05\columnwidth]{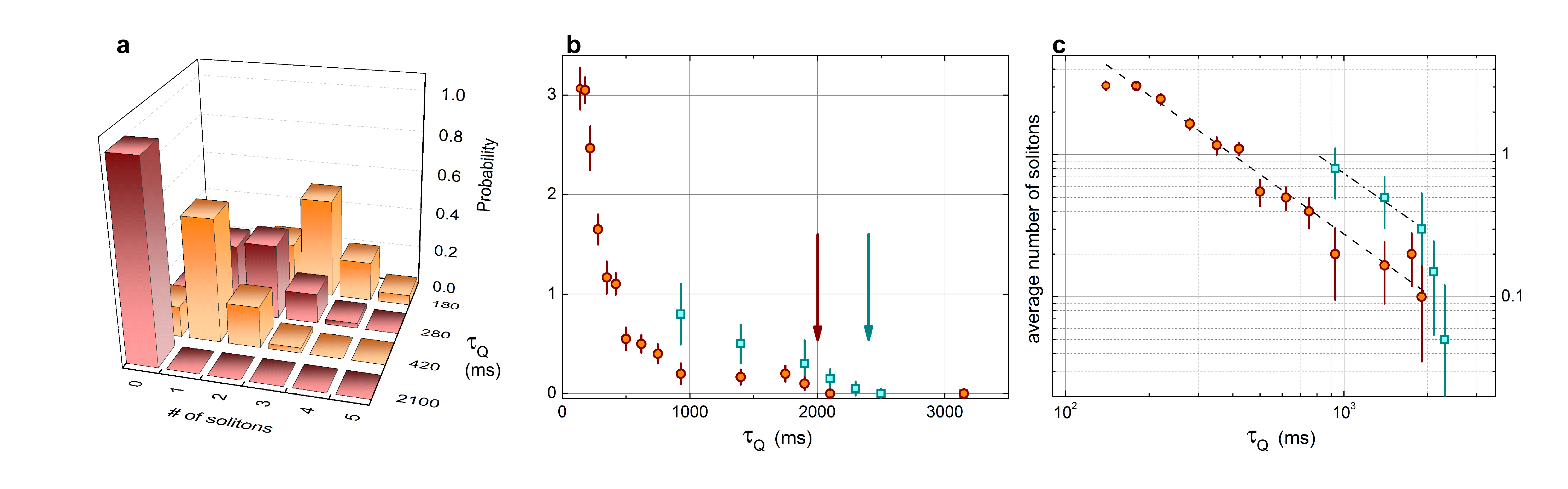}
\caption{\textbf{Soliton number vs. quench time:} \textbf{a}, Counting statistics of the number of solitons observed in each shot for four different quench times and for the data set with 25 million atoms at $T_\mathrm{c}$. Lin-lin (\textbf{b}) and log-log (\textbf{c}) plots of the average soliton number observed as a function of the quench time. Red circles and blue squares correspond to series of data taken with different number of atoms at the BEC transition, respectively 25 and 4 millions. Arrows in panel (\textbf{b}) indicate the maximum $\tau_\mathrm{Q}$ for which solitons were observed. The black dashed line in panel (\textbf{c}) shows the power-law dependence with exponent $-1.38\pm0.06$ as resulting from the best fit with the data points (red circles), excluding the point at the fastest quench. Dot-dashed line with the same slope, but shifted on the second data set, serves as a guide to the eye showing similar power-law. Reported error bars include the standard deviation on the average counts, and the resolution limit $1/N_\mathrm{meas}$ added in quadrature.}
\label{fig:statistics}
\end{figure*}

\begin{figure*}[!]
\centering
\includegraphics[width=2.05\columnwidth]{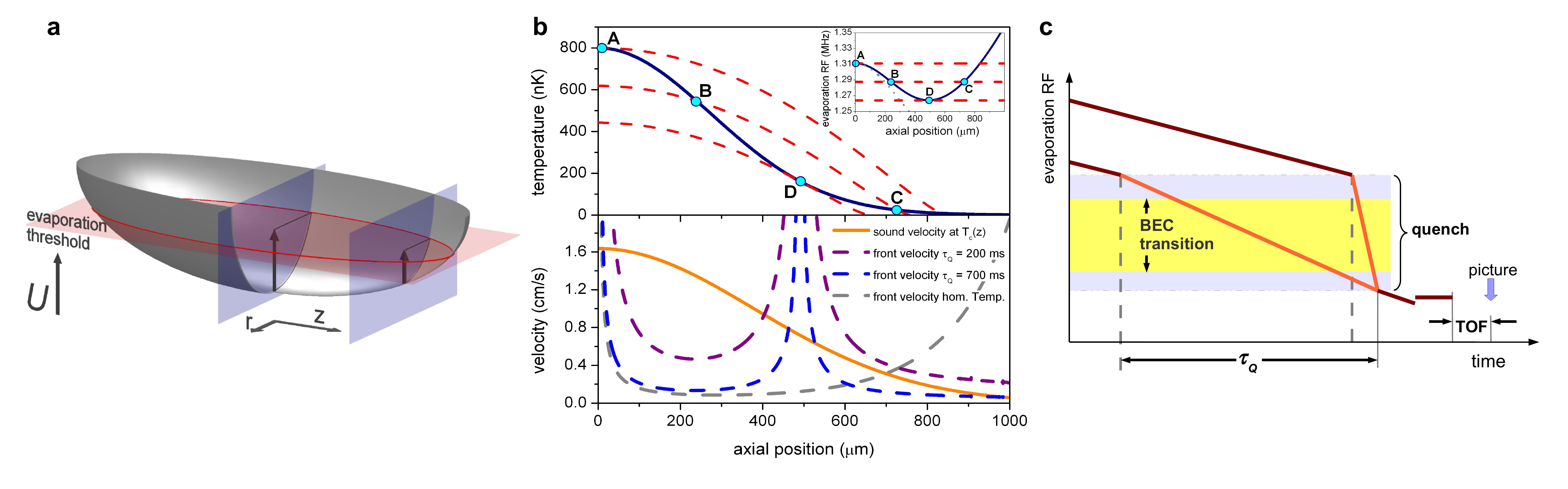}
\caption{\textbf{Quenched evaporation in an inhomogeneous trapped gases:} \textbf{a}, Sketch of the trapping potential showing the elongated weakly confined direction, $z$, and the radial tightly confined one, $r$,
and the evaporation threshold set by the RF field (horizontal plane). This is reduced quickly in time in such a way that, during the quench, atoms thermalize only radially, according to the local potential depth (see different vertical planes) giving rise to a local temperature $T(z)$. The inhomogeneous density distribution also causes a local critical temperature $T_\mathrm{c}(z)$. \textbf{b}, Top graph shows the profile of the critical temperature (solid blue line) and three different temperature profiles  (dashed red) corresponding to different evaporation thresholds. During quenched evaporation condensation is first reached at the trap center (A), then the critical point shifts towards outer regions (B) but a second front also enters from the edge (C), finally the transition is reached also at the last point (D). The inset shows the critical frequency along $z$. The lower graph shows the local sound velocity (solid orange), the local speed of the transition front for two values of the quench time, namely 700 ms (dashed blue) and 200 ms (dashed pink), and the front speed in the case of a uniform temperature profile (dashed grey). \textbf{c}, Experimental sequence: the first part of evaporation is always the same from a hot sample to a cold one above $T_\mathrm{c}$; then a ramp with variable quench time brings the system from above to below $T_\mathrm{c}$ (light blue regions); a short final cooling ramp allows to increase the condensate atom number and 100~ms are left to equilibrate the system in trap. Yellow frequency band corresponds to the extended region in which $T_\mathrm{c}$ is crossed in the system (see Methods). Atoms are then imaged after 180 ms of TOF.}
\label{fig:GeneralSketch}
\end{figure*}

A key point of our analysis is that the number of defects that we observe is larger when the quench is faster, as reported in Fig.\,\ref{fig:statistics}. This is a clear indication that our solitons are produced via the KZM. In order to provide a quantitative support to this scenario we need to check whether, for a given quench time, the transition front propagates faster than the causal horizon hence activating the KZM \cite{Zurek09}. To this aim, the details of the trapping potential and the evaporation procedure are relevant. Sodium atoms are trapped in an elongated magnetic potential, whose profile is sketched in Fig.\,\ref{fig:GeneralSketch}a (see Methods). The evaporation threshold is set by a radio-frequency $\nu_\mathrm{RF}$ tuned to flip the atomic spin, from the trapped to the untrapped state, at a given potential energy from the bottom of the trap. The effective evaporation threshold is governed by the radial motion of the atoms and depends on $z$, being fixed by the difference between the evaporation threshold at the trap bottom ($r=z=0$) and the local axial potential $U(r=0,z)$. Moreover, the elastic collisional rate is large enough to ensure local thermal equilibrium (collisional regime) but with a temperature gradient along the axial direction \cite{Comment:collision-hydrodinamic}. For these reasons, we define an axial temperature $T(z)$ equal to the corresponding evaporation threshold expressed in thermal units, divided by the truncation parameter $\eta$  which is of the order of $5$ in our case \cite{Ketterle96,Comment:temperatureVSgravity-sag}:
\begin{equation*}
T(z)=\frac{h\,\nu_\mathrm{RF}-U(r=0,z)}{\eta\,k_B}.
\end{equation*}
Typical temperature profiles for three values of evaporation radio-frequency are shown in the top panel of Fig.\,\ref{fig:GeneralSketch}b (red dashed lines). \\

The cooling process starts with a ramp of radio-frequency forced evaporation down to a temperature 10\,\% higher than the largest critical value for observing a condensate fraction in our sample (see Fig.\,\ref{fig:GeneralSketch}c). At this stage the gas is non condensed and in thermal equilibrium. We can estimate the profile of the critical temperature $T_c(z)$ by inserting the above-$T_c$ equilibrium density distribution of the cloud in the expression of $T_c$ for noninteracting particles:
\begin{equation*}
 T_\mathrm{c}(z)=\frac{2\pi\hbar^2}{mk_B}\left(\frac{n(r=0,z)}{\zeta(3/2)}\right)^{2/3}
\end{equation*}
where $m$ is the atom mass and $\zeta(...)$ the Riemann $\zeta$-function. A typical result is shown in the top panel of Fig.\,\ref{fig:GeneralSketch}b (solid blue line) for a sample of $25\times 10^6$ atoms. Then the system is thermally quenched by linearly reducing the evaporation threshold down to a value such that $T(z)< T_\mathrm{c}(z)$ everywhere. During this process, the local temperature profile crosses the local critical temperature profile at some values of $z$, which define the positions of the BEC planar transition fronts propagating along $z$ as the temperature lowers. The speed of the transition fronts depends on $z$ and on the quench time $\tau_\mathrm{Q}$. The latter can be varied by keeping the initial and final radio-frequencies fixed, but changing the duration of the evaporation process  (see Fig.\,\ref{fig:GeneralSketch}c and Methods). The evaporation ends with a final "slow" ramp followed by an equilibration time (both lasting 100 ms). \\

The speed of a transition front can be estimated from the curves of $T(z)$ and $T_\mathrm{c}(z)$, as those plotted in the top panel of Fig.\,\ref{fig:GeneralSketch}b. For the speed of the causal horizon, i.e. the fastest speed at which the information about the choice of a local macroscopic phase of a BEC can travel across the gas, we take the speed of sound $v_\mathrm{s}$  (sonic horizon). A precise determination of this quantity in the vicinity of the transition and for a non-uniform gas is highly nontrivial. As a reasonable estimate we can use the expression for the sound speed derived in Ref.\,\cite{Hu10} within a two-fluid model; near $T_\mathrm{c}$, it gives  $v_{\mathrm{s}}^2(T)=\frac{5\zeta(5/2)}{3\zeta(3/2)}k_BT+2gn$, where $g$ is the interaction parameter related to the s-wave scattering length $a$ by $g=4\pi \hbar^2 a/m$. The bottom panel in Fig.\,\ref{fig:GeneralSketch}b shows the comparison between the speed of the transition front for different quench times (dashed lines) and the local sound speed in the gas $v_\mathrm{s}(z)\propto \sqrt{T_\mathrm{c}(z)}$ when neglecting interactions (solid line). The figure shows that indeed there are regions, both near the center and in the tails of the atomic distribution, where the transition front moves faster than the sonic causal horizon and that the spatial extension of those regions depends on the quench time. \\

For given experimental conditions the number of defects we observe varies from shot to shot, as expected from the stochastic nature of the KZM. We do a quantitative characterization by counting the number of solitons observed over a large number of realizations $N_\mathrm{meas}$ (see Methods). The normalized statistic probability of detecting a given number of solitons, which we report in Fig.\,\ref{fig:statistics}a for $0$ to $5$ counts and four different quench times, follows the Poissonian distribution. In panels b) and c) of the same figure we plot, both in lin-lin and log-log scales, the average number of detected solitons as a function of  the quench time, by varying $\tau_\mathrm{Q}$ over more than one order of magnitude. Results are shown for two set of measurements done with a high (red circles, $N_{\rm at} = (25 \pm 5)\times 10^6$) and low (blue squares, $N_{\rm at} = (4 \pm 1) \times 10^6$) number of atoms at the transition. For each set, the vertical arrow in Fig.\,\ref{fig:statistics}b indicates the maximum quench time (i.e., minimum quench rate) for the observation of solitons; for larger values of $\tau_\mathrm{Q}$ solitons are never observed in our sample. For lower values of $\tau_\mathrm{Q}$ the average number of solitons exhibits a power-law dependence on the quench time as expected for the KZM.  For the largest condensate and for $\tau_\mathrm{Q}$ shorter than $140$~ms, the generation of solitons is accompanied by a marked loss of atoms at the end of the evaporation ramp hence resulting in a significant reduction of the radius of the final condensate. For smaller condensates this constraint in $\tau_\mathrm{Q}$ is stronger, reducing the accessible range for testing the KZM. The capability of producing large condensates is thus crucial for this type of experiments. \\

The dashed line in Fig.\,\ref{fig:statistics}c is the power-law $\tau_\mathrm{Q}^{-\alpha}$ with the exponent $\alpha=1.38\pm0.06$ obtained by fitting the experimental data with higher $N_\mathrm{at}$.  Also the second series, with the smaller condensate, seems to follow the same power-law dependence, thus confirming the universal nature of the mechanism. This result for the exponent can be compared with the prediction given by Zurek \cite{Zurek09} for the formation of gray solitons in a cigar-shaped condensate; depending on the choice of the values of the critical exponents for the coherence length and the relaxation time of a Bose gas at the BEC transition, the predicted value was $\alpha=1$ or $7/6$. The order of magnitude is the same, but the comparison should be taken with care. The calculation of Ref.\,\cite{Zurek09} assumed a uniform temperature in the gas, while in our experimental conditions (trapping frequencies and collisional rate) the local temperature $T(z)$ is non-uniform.  The difference can be appreciated by looking at the lower panel of Fig.\,\ref{fig:GeneralSketch}b, where the speed of the transition front for the case of a uniform temperature is shown as the lower gray dashed line and compared with the local speed of sound. As one can see, the front speed is larger than the sound speed only in a narrow region near the center of the atomic distribution, where the defects can nucleate at the transition (in the outer part of the cloud, the front is also faster than sound, but the density is vanishingly small). Conversely, with a non-uniform temperature profile, solitons can form also in a region in the tail of the distribution, where the density is still large enough (around point D in the upper panel); this region becomes wider as the quench time is reduced. The fact that defects can nucleate in different regions of the gas favors the observation of a larger number of solitons; for the same reason, the observed value of $\alpha$ may be different from the value predicted in \cite{Zurek09}. \\

A further argument supporting the interpretation of our observations in terms of the KZM is how the threshold for detecting solitons varies by varying the atom number in the gas. The two sets of data in Fig.\,\ref{fig:statistics}b have different atom numbers. Using the expression by Hu \emph{et al.} \cite{Hu10} for the sound speed, near $T_\mathrm{c}$ one gets approximately $v_\mathrm{s}\propto \sqrt{T_\mathrm{c}}\propto N_\mathrm{at}^{1/6}$. Thus in the case of lower atom number the sound speed at $T_\mathrm{c}$ is smaller by a factor $0.74\pm0.06$. If the sound speed is lower, defects can be created for smaller quench rates (larger quench times), as we indeed observe.  The positions of the two arrows in Fig.\,\ref{fig:statistics}b differ by a factor $0.84\pm0.08$, in reasonable agreement with the expectations of the above illustrated model. \\

\begin{figure}[!]
\centering
\includegraphics[width=0.8\columnwidth]{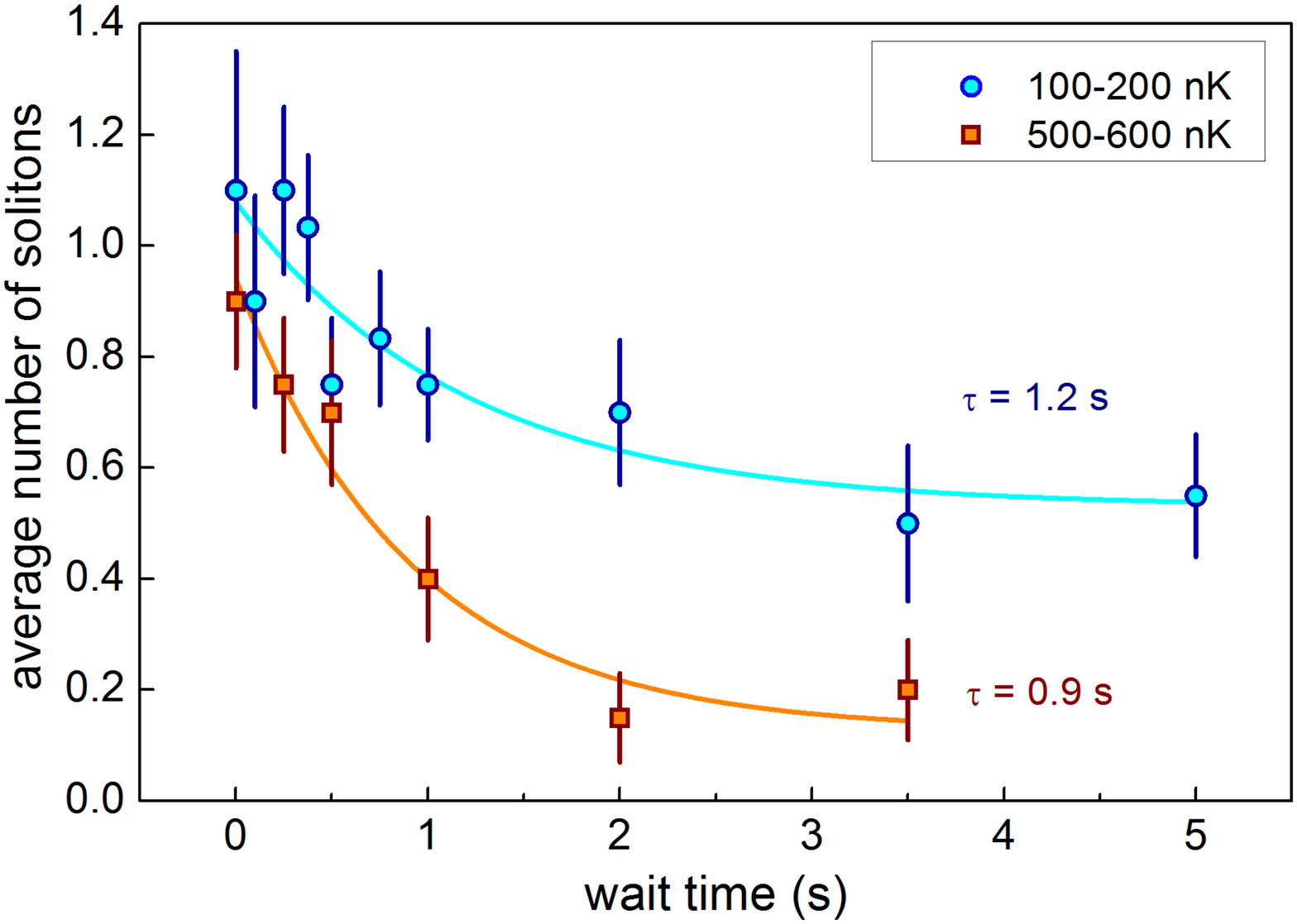}
\caption{\textbf{Soliton number decay:} blue circles correspond to a very cold sample, with a barely visible thermal fraction ($T\simeq100\div200$ nK), whereas red squares were acquired for a hotter sample, close to $T_\mathrm{c}$ ($T\simeq500\div600$ nK). Solid lines show the exponential fit to the decay. $\tau_\mathrm{Q} = 467$ ms for both datasets.}
\label{fig:Lifetime}
\end{figure}

We finally discuss the possible spurious effects that may alter the counting of solitons, hence modifying the data reported in Fig.\,\ref{fig:statistics}. We already mentioned that, in the limit of very fast quenches, the efficiency of the evaporative cooling drops causing additional losses of atoms at the transition and in the final condensate. The point at $\tau_\mathrm{Q} = 140$ ms lies in a range where these effects may start being relevant.  On the other side, when a slow quench is applied, one may wonder if solitons decay before being observed. With this regard in Fig.\,\ref{fig:Lifetime} we report the average number of solitons observed in the condensate as a function of the time passed in-trap after an intermediate evaporation quench time. Two sets of data are collected for different values of the temperature after the evaporation. In both sets, the number of solitons shows a decay with a tendency to saturate to a finite value for long times. The decay is faster when the temperature is larger (slightly smaller than $T_c$), consistently with the idea that, at finite $T$, gray solitons are accelerated towards the edges of the cloud where they can dissipate their energy into thermal excitations.  If $T\ll T_c$, in the worst case of a wait time of 2 seconds, the counts drop only by about 30\,\%, which is a reasonably small loss when compared to the error bars of Fig.\,\ref{fig:statistics}. The tendency to saturate at a finite value suggests that a sizable fraction of solitons has a much longer lifetime. This can be explained in terms of the generation of dark solitons near the center of the trap. As opposed to gray solitons, which travel back and forth along the BEC axial direction probing regions with lower quantum degeneracy, dark solitons experience  the lowest temperature (highest quantum degeneracy) available in the sample, and hence are subject to a slower decay. \\

To summarize, we report on the spontaneous nucleation of solitons in a BEC of sodium atoms via the inhomogeneous Kibble-Zurek mechanism by crossing the phase transition at a finite rate. The relatively large atom number at the transition, together with the evaporation occurring in a regime of local thermal equilibrium with a non-uniform temperature profile, allows us to observe up to 5-6 solitons in a single sample, to extract a power-law dependence of the average number of solitons on the quench time, and to provide a check of the KZM scaling with the sonic horizon. In combination with the observations of \cite{Weiler08}, our results should stimulate the investigation of the interplay between the inhomogeneous and homogeneous KZM \cite{delCampo11}. In addition, an extension of the theory of Ref. \cite{Zurek09} to the case of non-uniform temperature profiles could allow to extract from our observations the values of the critical exponents for Bose-Einstein condensation in dilute gases.\\

\bigskip

\textbf{METHODS SUMMARY}\\
Sodium atoms, collected from a compact high-flux source \cite{Lamporesi13}, are evaporatively cooled in a cigar-shaped Ioffe-Pritchard magnetic trap \cite{Pritchard83} with harmonic trap frequencies $\omega_\mathrm{z}= 2\pi \times 12$ Hz and $\omega_\mathrm{r}= 2\pi \times 119$ Hz. The duration of a part of the evaporation ramp across the transition is varied in order to explore different quench rates. Atoms are then released from the trap and let expand, but preventing their fall by levitating them with a vertical magnetic field gradient. After 180\,ms the condensate distribution has a pancake shape and its optical density is low enough to be imaged from any radial direction without saturation. This allows to clearly detect the presence of solitons that are formed in trap during the quench.
Further details are provided in the full Methods.

\bigskip

\textbf{Acknowledgements} We are indebted to L. P. Pitaevskii, I. Carusotto and A. Recati for fruitful discussions. This work is supported by Provincia Autonoma di Trento.

\bigskip

\textbf{Author Contributions} G.L, S.D., S.S. and G.F built the experimental set-up; G.L, S.D. and G.F performed data acquisition; G.L. and G.F. analyzed the data; all authors contributed to the discussion of the results and G.L., S.D., F.D. and G.F. participated in manuscript preparation.

\bigskip

\textbf{Author Information} The authors declare no competing financial interests.

\bigskip

\textbf{METHODS}\\
\textbf{Sample preparation.}\
A high-flux beam of cold sodium atoms is produced in a compact system based on a short Zeeman slowing stage and on a coplanar 2D MOT \cite{Lamporesi13}. The bright atomic beam fills a 3D dark spot MOT in 8 seconds. A few ms dark spot molasses helps increasing the phase space density to $3\times 10^{-6}$ and improve the transfer efficiency into the magnetic trap. Our magnetic trap has a Ioffe-Pritchard \cite{Pritchard83} geometry with final axial and radial trapping frequencies of $\omega_\mathrm{z}= 2\pi \times 12$ Hz and $\omega_\mathrm{r}= 2\pi \times 119$ Hz. Starting with $10^9$ atoms in the magnetic trap we cool them by means of a Zeeman forced evaporation with a two-step ramp, first reducing the radio-frequency at 1.2 MHz/s for 30 seconds, then the trap is decompressed by a factor $\sqrt{2}$ to the final trapping frequencies and evaporation continues at 190 kHz/s for 8 seconds. At the end of the preparation stage the atomic sample contains $25\times 10^6$ atoms just above $T_\mathrm{c}$. The trapping potential has a cigar shape horizontally oriented along $z$.\\
\textbf{Temperature quench.}\
As illustrated in Fig.\,\ref{fig:GeneralSketch}c we explore the KZM by crossing the BEC transition point with different evaporation quench rates. The starting point is always 1.39 MHz, 190 kHz above the trap bottom. The BEC transition frequency is not unique. In order to make sure to always cross the BEC transition throughout the whole sample during the quench we need to take several effects into account: a) the density inhomogeneity across the sample introduces a transition frequency interval, b) changing the quench rate slightly shifts the transition point because of the different amount of removed atoms, c) technical shot-to-shot atom number fluctuation also shifts the transition point. For all these reasons we set a fixed frequency band from 1.39 MHz to 1.25 MHz, within which the whole sample crosses the BEC transition for any given experimental quench rate. The quench time $\tau_\mathrm{Q}$ reported in the text is defined as the time interval employed to perform this linear quench ramp of 140 kHz. The quench is followed by a 100 ms long further evaporation at 300 kHz/s down to 1.22 MHz, in order to maximize the condensate fraction, and a final 100 ms during which the RF is kept fixed at 1.24 MHz allowing for solitons stabilization and evolution.\\
\textbf{Levitation.}\
Switching off the elongated magnetic trapping potential the chemical potential is suddenly transferred into kinetic energy and atoms mainly expand along the tightly confined radial direction assuming a spherical shape after about 15 ms, then the atomic distribution becomes pancake-like. In the meantime the sample would naturally fall under gravity and reach the glass vacuum cell within 50 ms time. To avoid this and allow for longer expansion times in order to observe atoms without optical density saturation, we levitate the sample by switching on just one of the two quadrupole coils producing a vertical gradient of the magnetic field modulus, able to compensate the gravitational force for atoms in $\Ket{F=1,m_\mathrm{F}=-1}$. The residual magnetic field curvature in the horizontal plane gives rise to a negligible trapping effect.\\
\textbf{Imaging.}\
The least energetic and most stable orientation for solitons in an elongated BEC is the one orthogonal to the trap symmetry axis. Density depletion can be therefore observed by looking along any radial direction. We image the condensates along two orthogonal directions in the radial plane in order to minimize underestimating soliton counts due to any possible residual tilt of the solitonic plane. Absorption imaging is performed after a levitation time of 180 ms. In this way the optical density of the condensate is of the order of 1 or smaller, no saturation is present and density variations in the sample are clearly visible. Atoms are imaged using light resonant with the $\Ket{F=2}\rightarrow\Ket{F'=3}$ transition. Since atoms are magnetically trapped in $\Ket{F=1,m_\mathrm{F}=-1}$, repumping light tuned on the $\Ket{F=1}\rightarrow\Ket{F'=2}$ transition is needed to pump them in $\Ket{F=2}$. A thin light-sheet (waist of 600 $\mu$m) propagating along the vertical direction, is used to repump only a central region of the expanded condensate (Thomas-Fermi radius of 2.5 mm) in order to further reduce optical density and increase the soliton contrast.\\
\textbf{Data analysis.}\
For each set of experimental parameters, such as temperature, quench time and atom number at the transition, the experiment was repeated for 20 to 40 times (depending on the resulting average number of solitons observed) in order to minimize the error bars in Fig.\,\ref{fig:statistics}. The number of solitons visible in each image was counted (see Fig.\,\ref{fig:solitons}b-e) and the average number was plotted for any given set of parameters. Error bars include the standard deviation on the average counts, and the resolution limit $1/N_\mathrm{meas}$ added in quadrature.

\bigskip

\noindent


\begin{thebibliography}{110}
\bibitem{Comment:collision-hydrodinamic}
\bibinfo{title}{{\rm The axial collisional regime follows from a collisional
  rate tipically larger than the axial trapping frequency. For insance at the
  trap center the collisional time is of the order of 5 ms, to be compared with
  the axial oscillation period of 83 ms.}}
\bibitem{Comment:temperatureVSgravity-sag}
\bibinfo{title}{{\rm Minor corrections can also be included in order to account for effects of gravity on the confinining potential.}}
\end{thebibliography}
\end{document}